\documentclass[12pt]{article}
\usepackage{epsfig}

\newlength{\dinwidth}                                                   
\newlength{\dinmargin}                                                  
\begin{document}
%
%
%
\thispagestyle{empty}
\setlength{\unitlength}{1mm} 
\begin{center}                                                         
\large 
Work is supported by Grant of INTAS-RFBR, Project No  95-0679 \\
\normalsize
\end{center}
%
%
\vspace*{40mm}
\begin{quotation}
\renewcommand{\baselinestretch}{1.0}\large\normalsize  
\begin{center}
\Large {\bf A New High Energy Photon\\
       Tagger for the H1-Detector at HERA} 
\end{center}
\vspace*{5mm}
\normalsize
 \begin{center} 
{\it   V.F. Andreev$^a$, A.S. Belousov$^a$, A.M. Fomenko$^a$, L.A. Gorbov$^a$\\
     T. Greenshaw$^b$, S.V. Levonian$^a$, E.I. Malinovski$^a$, S.J. Maxfield$^b$\\
    I.P. Sheviakov$^a$, P.A. Smirnov$^a$, Yu.V. Soloviev$^a$, G.-G. Winter$^c$ } 
 \end{center}
\vspace*{20mm}    
 \begin{center}
{\bf Abstract}  
 \end{center} 

\small
\noindent 
The H1 detector at HERA has been upgraded by the addition of 
a new electromagnetic calorimeter. This is installed in the 
HERA tunnel close to the electron beam line at a position 
$ \simeq 8\,$m from 
the interaction point in the electron beam direction.
The new calorimeter extends the acceptance
for tagged photoproduction events to the high $y$ range,
$0.85 < y < 0.95$, and thus significantly improves the capability
of H1 to study high energy $\gamma$-p processes. 
The calorimeter design, performance and first results obtained 
during the 1996-1999 HERA running are described.

\renewcommand{\baselinestretch}{1.2}\large\normalsize
\end{quotation}
\vspace*{40mm}
------------------------------------------------------ \\
$^a$  - {\it Lebedev Physical Institute, Moscow, Russia }\\
$^b$ - {\it Deparment of Physics, University of Liverpool, Liverpool,
   UK} \\ 
$^c$ - {\it DESY, Hamburg, Germany}         \\                                                        
%
\newpage  
\section{Introduction}
The HERA accelerator at DESY brings into 
collision electrons (or positrons) with energy $27.5\,$GeV
and protons with energy $920\,$GeV ($820\,$GeV prior to 1998).
A large proportion of the resulting interactions can be considered
to occur between protons and essentially real
photons, that is, photons for which $Q^2 \sim 0$, where $Q^2 = -q^2$ and $q$ is the four-momentum 
of the photon exchanged between the electron and 
the proton. 
The energy, $E_{\gamma}$, of the photons
in these photoproduction interactions
can be related to the energies $E_e$ and $E'_e$ of the electron
in the initial and
final states through the expression
$$
y=1-\frac{E'_e}{E_e}+\frac{Q^2}{4E_e^2} 
 \simeq 1-\frac{E'_e}{E_e}
 =\frac{E_{\gamma}}{E_e}.
$$
Here $y$ is defined by
$y=p \cdot q/p \cdot k$, $p$ and $k$ being the four-vectors
of the initial proton and electron, respectively.
Measuring the energy of the outgoing electron enables the
deduction of the photon's energy; the photon is then 
said to have been ``tagged''.  
In order to study the highest energy photon-proton collisons possible,
a new electromagnetic calorimeter has been added to the 
H1 detector at HERA. This extends the range over which 
photons can be tagged from
$0.3 < y < 0.8$ and  $0.08 < y < 0.15$, using the   
electron taggers ET and ET44, respectively, to include the 
region $0.85 < y < 0.95$. In terms of the accessible 
photon-proton centre-of-mass energies, these regions correspond to 
$170 < W_{\gamma p} < 280\,$GeV, 
$90 < W_{\gamma p} < 120\,$GeV and
$290 < W_{\gamma p} < 310\,$GeV, respectively.
  
The electron taggers ET (or ET33) and ET44 are
electromagnetic calorimeters located
$33.4\,$m and $44\,$m from the interaction point (IP) in the 
electron direction, the negative $z$ direction
in the H1 co-ordinate system.
The new calorimeter, the ET-8, 
is located at $z \simeq -8\,$m from the IP to enable the measurement 
of electrons scattered with 
low energies while satisfying the constraints arising from the HERA 
machine: the space available for the ET-8 is limited and the 
environment hostile. 
This paper describes how
the design of the ET-8 
allows the triggering on and the measurement of 
low energy electrons with 
the necessary accuracy and ensures that it is sufficently
radiation hard to survive operation at HERA.
Test results are presented, as are the results of first 
physics studies using the ET-8.
                                                          
\section{Realization}
\subsection{Layout of H1 Lumi system} 

This technically challenging project requires the detection
and accurate measurement of low energy electrons, down 
to 1.5 GeV, in a high radiation environment. The ET-8
must thus be radiation hard and have good energy resolution.
In addition, its response to hadrons must be well understood
to facilitate the separation of the signals due to electrons
from those due to hadrons. Large hadronic background is expected
due to the proton beam halo.
 \begin{figure}[ht]
 \centering
 \begin{picture}(160,70)(0,0)
  \epsfig{file=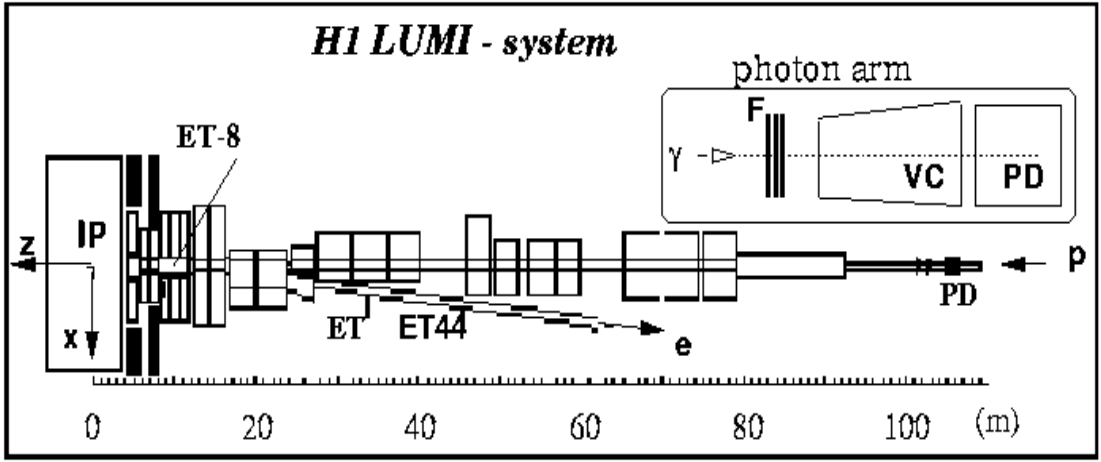,height=70mm,width=150mm,%
   bbllx=-40pt,bblly=260pt,bburx=640pt,bbury=530pt,clip=,angle=0}
\end{picture}
\caption{Schematic top view of the H1 luminosity system.}
\label{fig:lumi}
\end{figure}
 
A plan view of the H1 luminosity system is shown in 
figure~\ref{fig:lumi}.
Visible are the abovementioned electron taggers, 
the ET and ET44, and the photon detector (PD).
Electrons and photons originating from
bremsstrahlung processes at the interaction point (IP) are emitted
at very small angles and have to be detected
in the luminosity system. The axis of the photon beam coincides
with the axis of the incident electron beam which,
under nominal operation conditions, 
coincides with the $z$-axis of H1. 
The bremsstrahlung photons propagating inside the 
proton beampipe leave it through an exit window at $z = -92\,$m
and hit the photon arm of the luminosity system, of which the main 
part is the PD. The lead filter (F) and a water
\v Cerenkov counter (VC) are located in front of the PD. The filter,
of $2\,X_0$ thickness, 
absorbs most of the synchrotron radiation flux and
protects the PD against radiation damage.
At nominal luminosity, the
direct synchrotron radiation flux carries a power of about 
$0.45\,$kW.
The water VC acts as 
an additional $1\, X_0$ long active filter and can reject events
in which a photon has converted  between the exit window and
the photon detector. It also provides an energy measurement for the
absorped part of electromagnetic showers. 

Accurate determinations of the luminosity are made 
using measurements of the e-$\gamma$ coincidence
rate within the 
acceptance of the luminosity system. 
Rapid luminosity monitoring during HERA 
running is provided by measurements of the rate of photons
striking the PD, the so-called ``single photon'' method.   
The scattered electrons are detected in the electron taggers ET-33 
and ET-44.
These are total absorbtion \v Cerenkov calorimeters made of 
KRS-15 crystals~\cite{LPI2}.
The new detector was incorporated into the H1 trigger system 
and operated during all HERA luminosity running to allow
the tagging of photoproduction events close to 
the HERA kinematical limit.
The location with 
respect to the electron and proton beams and the dimensions
of the detector must  be accurately known in order to determine 
its acceptance  with sufficient precision. 
An additional difficulty
is that the space 
available for the ET-8 between the  HERA 
beamline elements is very restricted. 
\subsection{Acceptance}

Monte Carlo simulations for acceptance studies were 
performed using a fast
simulation package H1LUMI~\cite{LEVO}.
This models the current HERA electron ring optics
(the proposed spin rotators have yet to
be included) as is described e.g. in~\cite{BRINK} .
Several possible positions for the new tagger 
were considered. 
%
\begin{figure}[h]
\centering
\begin{picture}(140,90)(-5,-2)
\epsfig{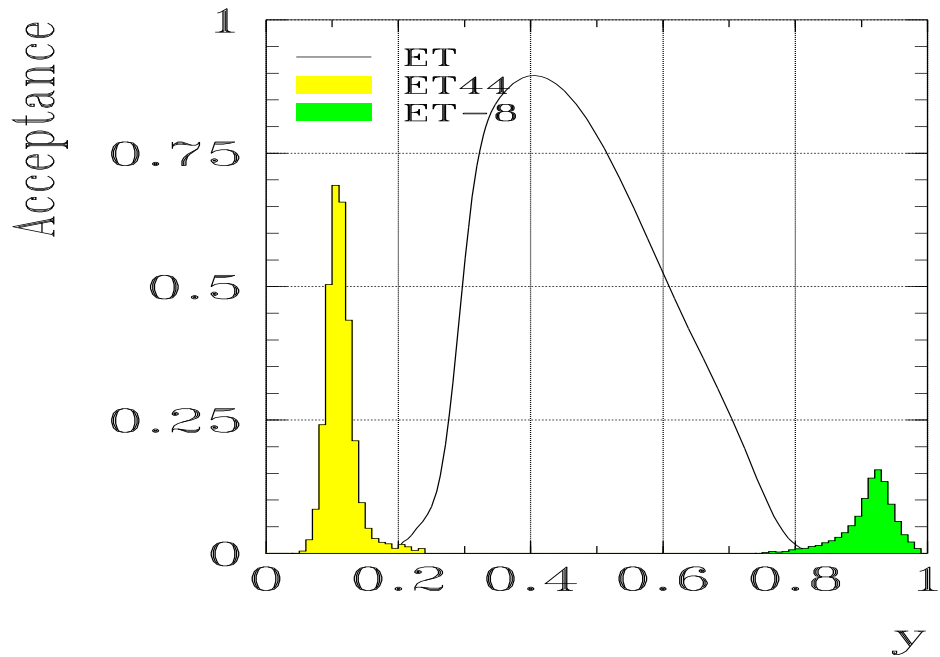}
\end{picture}
\caption{Acceptances of the three electron taggers for photoproduction
with nominal electron beam conditions.}
\label{fig:accept}
\end{figure}
After taking into account the
apparatus currently present in the HERA tunnel,
the space needed for
the calorimeter itself and the access requirements for
installation and maintenence,
it was decided to install the new  tagger
at the position $z \simeq -8\,$m. Small possible variations
of the beam parameters at the IP, horizontal and vertical tilts
of $\pm 0.15\,$mrad and offsets of $\pm 1\,$mm, were taken into
account in the simulation. These typically lead to $15$ to $25\%$
changes in the ET-8 acceptance.
Precise knowledge of the tagger position with respect to
the electron beam is thus very important.
The acceptance can
be defined with high precision from the data using 
$ep \rightarrow e \gamma p$ events, as is done for
the present ET  and ET44 calorimeters. This reduces
uncertainties related to the precision with which real beam
conditions can be simulated.
\begin{figure}[ht]
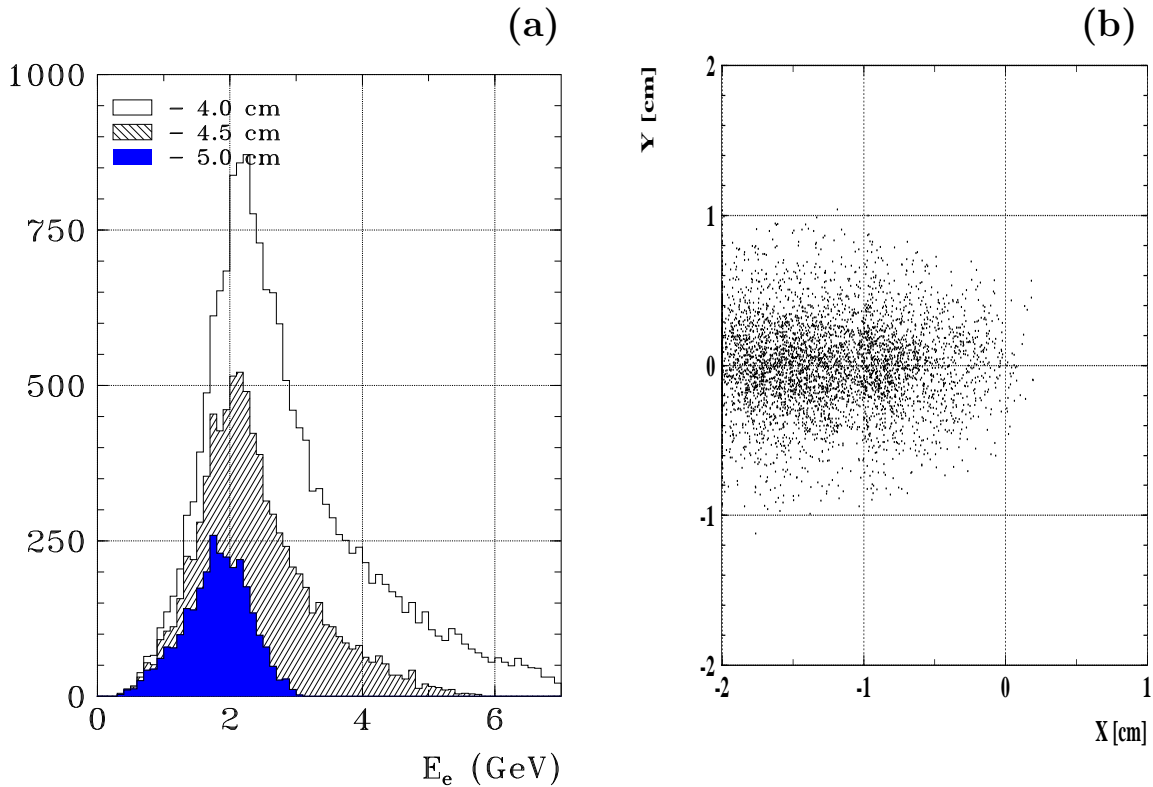

\centering
\begin{picture}(85,110)(0,1)
\put(70,100){\large {\bf (a)}}
\epsfig{file=fig4a.eps,width=80mm,height=100mm,%
   bbllx=20pt,bblly=2pt,bburx=540pt,bbury=660pt,clip=,angle=0}
\end{picture}
\begin{picture}(70,110)(0,1)
\put(60,100){\large {\bf (b)}}
\epsfig{file=fig4b.eps,width=70mm,height=100mm,%
 bbllx=60pt,bblly=384pt,bburx=515pt,bbury=770pt,clip=,angle=0}
\end{picture}
\caption{Energy and position distributions of the scattered 
electrons entering the ET-8:
   (a) energy distribution of the detected particles for various 
       ET-8 e-beam axis separations;
   (b) distribution of co-ordinates of point at which electron
       enteres the front side of the ET-8.}
\label{fig:Ehit}
\end{figure}
%
Figure~\ref{fig:accept} shows the acceptances of the 
ET, ET44 and ET-8 for
photoproduction with nominal beam conditions (zero tilt and 
offset). The ET-8 can be seen to extend the 
measurable $y$ range to the high $y$ interval
$0.85 < y < 0.95$. The acceptance of the calorimeter depends
on its position with respect to the electron beam axis, 
being most sensitive
to the distance between the beam and the 
edge of the active calorimeter
volume. The ET-8 acceptance shown here was calculated
using the nominal separation of 4.5 cm.

The energy distributions of the electrons accepted by the ET-8
at various distances from the electron beam axis
are shown in figure~\ref{fig:Ehit}(a). The distributions 
are those for 
photoproduction with $Q^2 < 10^{-2}\,$GeV$^2$.

The lateral size of the calorimeter was chosen on the basis
of the transverse distribution of the accepted electrons at 
the front plane of the detector. This distribution is shown
in figure~\ref{fig:Ehit}(b). It is seen that the majority of events 
have entry points
near the detector boundary close to the electron beampipe.  
Figure~\ref{fig:Ehit} illustrates that precise co-ordinate measurement
is necessary for good reconstruction of the electron energy,
allowing correction for those parts  
of the electromagnetic shower which leak out of 
the detector.
\subsection{Detector design}                                      
  
Various technical solutions were investigated before 
a final decision was made on the calorimeter 
type. Possible choices included, total absorption
calorimeters using KRS, NBW or PWO crystals,  a
lead-scintillator sampling calorimeter
and a spaghetti type calorimeter. Although the combination
of the total absorption crystal calorimeters and 
lead-scintillator calorimeters used in the existing H1
luminosity system, developed by the LPI group~\cite{H1KOL},
has proved to be 
very successful, it was decided that a spaghetti calorimeter,
as developed by H1 at DESY~\cite{H1SPA} 
was more suitable for the ET-8 detector. The SPACAL type
calorimeters are built from arrays of scintillating fibres embedded in
a heavy material, usually lead. The fibres inside these calorimeters 
transmit an amount of light which is proportional to the absorbed electromagnetic shower
energy  to a photo-sensitive readout device, for instance
a photomultiplier (PM). If the transverse dimensions of the shower
are considerably larger than the individual calorimeter cells, 
reasonably accurate determination of the shower position is 
possible.     
%
\begin{figure}[h]
\centering
\begin{picture}(150,200)(2,0)
\epsfig{file=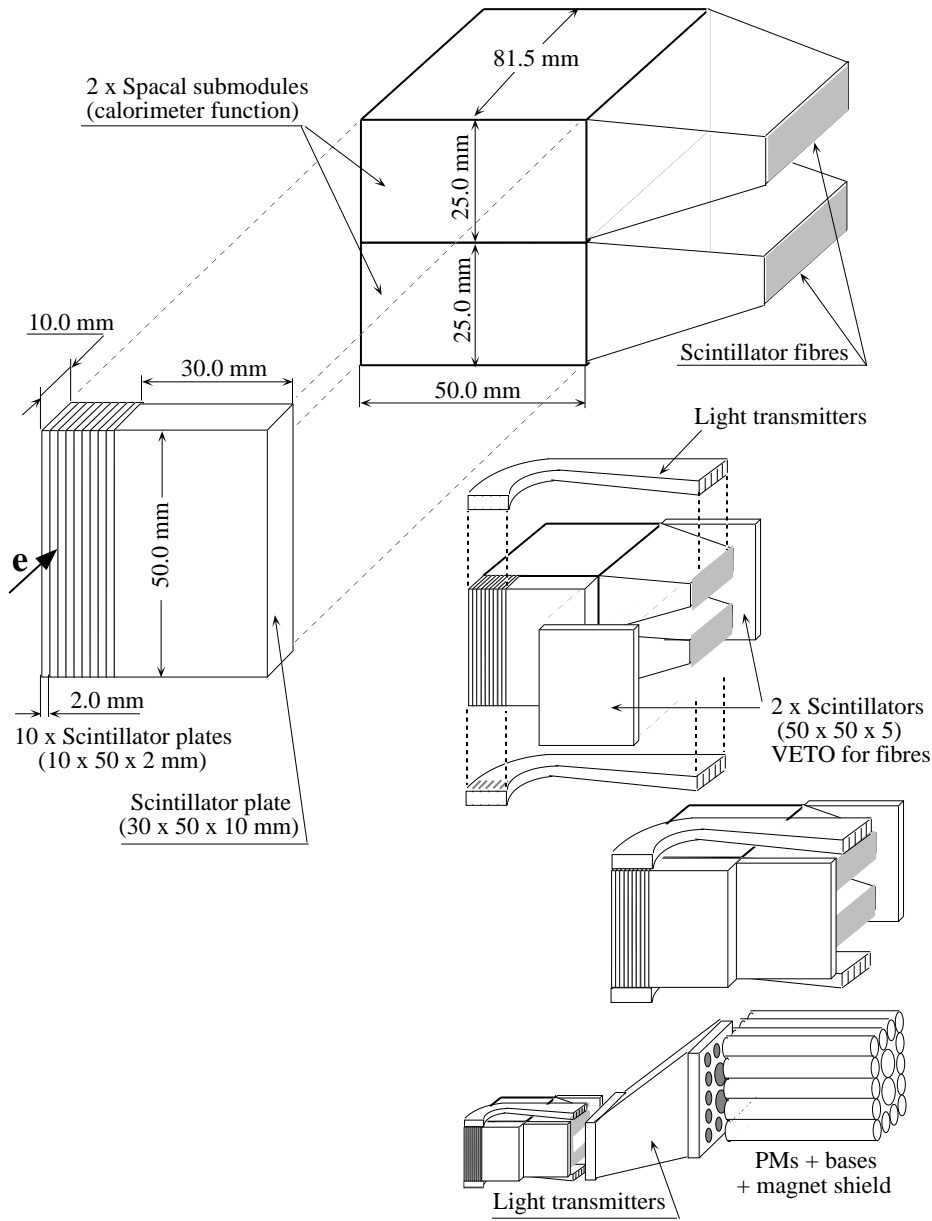,height=200mm,%
    bbllx=0pt,bblly=0pt,bburx=550pt,bbury=750pt,clip=,angle=0}
\end{picture}
\caption{Design of prototype ET-8 (the ET-7). 
      }
\label{fig:et7}
\end{figure}
%
BICRON BCF-12 fibres of $0.5\,$mm diameter, which emit blue light
with an emission peak near 430 nm, 
were used with a lead to fibre ratio of
2.3/1 by volume, giving a Moliere radius of $25.5\,$mm. 
This spaghetti type calorimeter has $\sim 1.5$ times better energy
resolution than the total absorption crystal calorimeters
mentioned above
and the combination with a scintillator hodoscope provides
the necessary spatial resolution. 

A prototype of the ET-8 calorimeter, called the ET-7 as 
it was positioned about $7\,$m from the interaction point,
was successfully tested during the July-October 1996 HERA running.
Following these tests, 
the decision was taken to leave the ET-7 in place and 
it is now a permanent
part of the H1 detector. Its design is shown in figure~\ref{fig:et7}. 
The active volume of
the ET-7 is $85\,$mm deep, corresponding to
$10\,X_0$. 
The necessary horizontal spatial resolution is provided by  the 
scintillator hodoscope placed in front of the calorimeter which allows
measurement of the co-ordinate of the entry position 
of the electron
with an accuracy of $1\,$mm using scintillator
plates of thickness $2\,$mm. The vertical co-ordinate 
is determinated from the ratio
of energies deposited in the two ET-7 modules.
\begin{figure}[h]
\centering
\begin{picture}(160,200)(-5,-7)
\epsfig{file=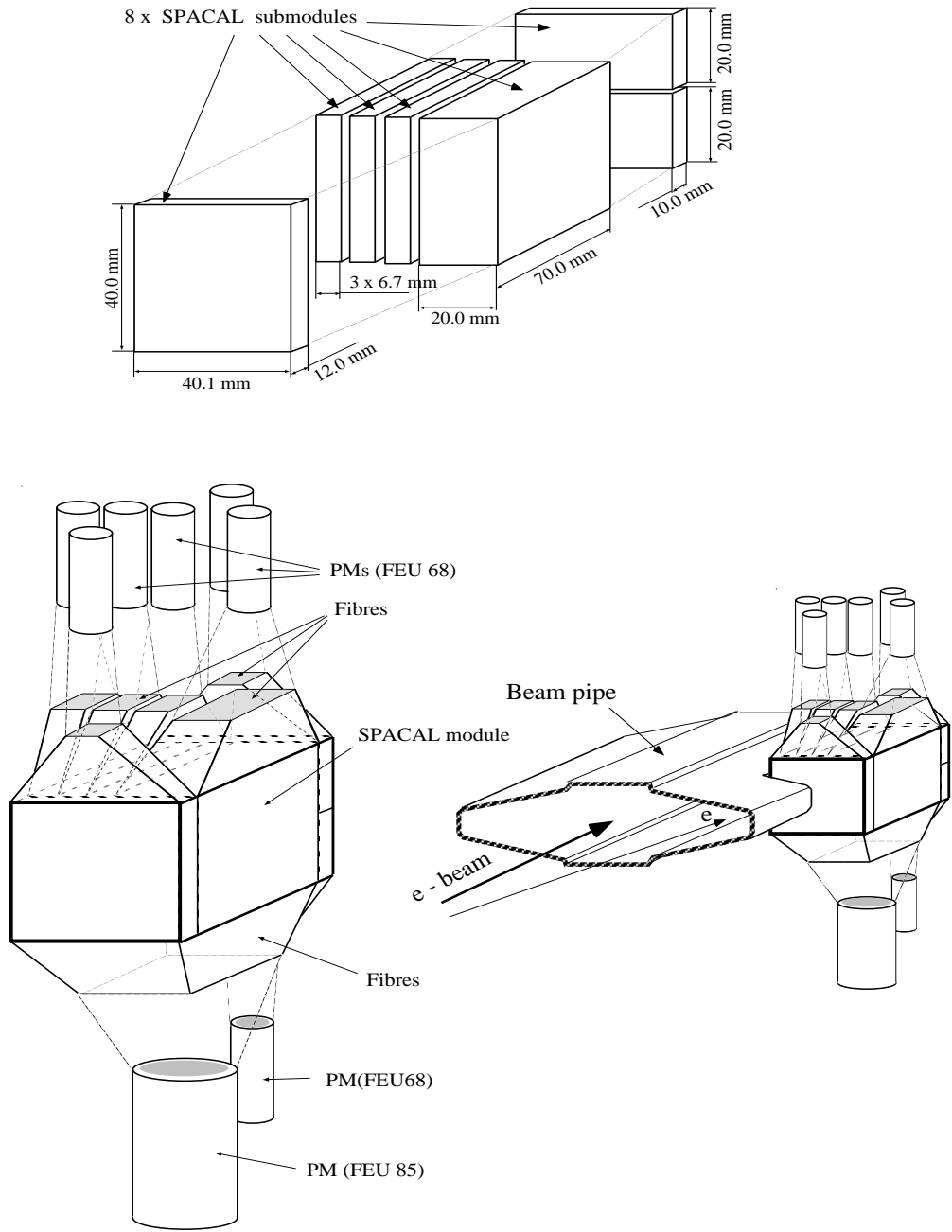,height=190mm,width=150mm,%
    bbllx=30pt,bblly=130pt,bburx=510pt,bbury=810pt,clip=,angle=0}
\end{picture}
\caption{Design of the ET-8 calorimeter
         which was installed in the HERA tunnel in 1997 }
\label{fig:et8}
\end{figure}
%
Two scintillator veto plates placed in front of and
behind the scintillator fibres are used to provide
background rejection.
The amount of  \v Cerenkov light from background 
particles in the light transmitters is orders of magnitude smaller
than the signal from electromagnetic showers in the calorimeter
due to the small lateral size of the light transmitters.
Hence it is not necessary to shield these with 
additional veto plates.
The detector was installed on a fixed support and  
stays permanently in the HERA median plane.

A schematic view of the ET-8 fibre detector, is shown 
in figure~\ref{fig:et8}.  
The total longitudinal
dimension of the active calorimeter volume 
is $92\,$mm.
The calorimeter 
consists of three parts, the front tagging counter made as a spacal
plate of $1.2\,X_0$ thickness, a central part and a rear co-ordinate
plane. The central part, with cross section $40 \times 40\,$cm$^2$ 
and of $70\,$mm
length, corresponding to $9\,X_0$, consists of 4 modules. 
As in the majority 
of accepted events the electron hits the ET-8  
close to the electron beam, the $3$ modules closest to the 
beamline have a
thickness of $6.7\,$mm, the module furthest away is thicker,
thus optimising the accuracy with which the 
horizontal co-ordinate of the electron's entry point 
may be found. This is done by  
by comparing the energy deposited in the modules in
the central part of the detector. 
The vertical co-ordinate is
determined from the ratio of output responses of the upper and
lower modules at the rear of the ET-8.
Showers caused by particles which did not enter through  
the front of the detector are 
vetoed by requiring coincidences with signals from the front counter.
  
The level 1 trigger element for the new tagger
requires the condition $E_{dep}^{tot} > E_{thr}$, where
$E_{dep}^{tot}$ is the total deposited energy in the
calorimeter and $E_{thr}$ the threshold energy which must 
have a low value, $E_{thr} \simeq 0.5-1.0$ GeV. For comparison
purposes, a minimum ionizing particle deposits 50 MeV in the ET-8.
An additional $\gamma$-veto condition from the photon 
arm of the luminosity system,
no signal in the PD, suppresses the 
background from the high rate Bethe-Heitler process, 
$ep \rightarrow e \gamma p $, with an efficiency of $\sim 98\%$.
%
\subsection{Radiation hardness of materials}  

The PD in the luminosity system is irradiated
by the bremsstrahlung beam from electron-proton
collisions and received a 
dose of $\approx 10\,$Mrad during the 1995 HERA running period 
and other elements of the luminosity system also receive large 
radiation doses.
The study of the radiation hardness of the materials used in the
construction of the ET-8, the 
scintillators, fibres and plexiglass, 
is thus an important task. These studies 
were carried out at
the $1\,$GeV electron synchrotron of the Lebedev Physical 
Institute in Moscow, Russia.
\begin{figure}[ht]
\centering
\begin{picture}(175,100)(0,0)
\put(155,85){\Large \bf (a)}
\put(155,35){\Large \bf (b)}
\epsfig{file=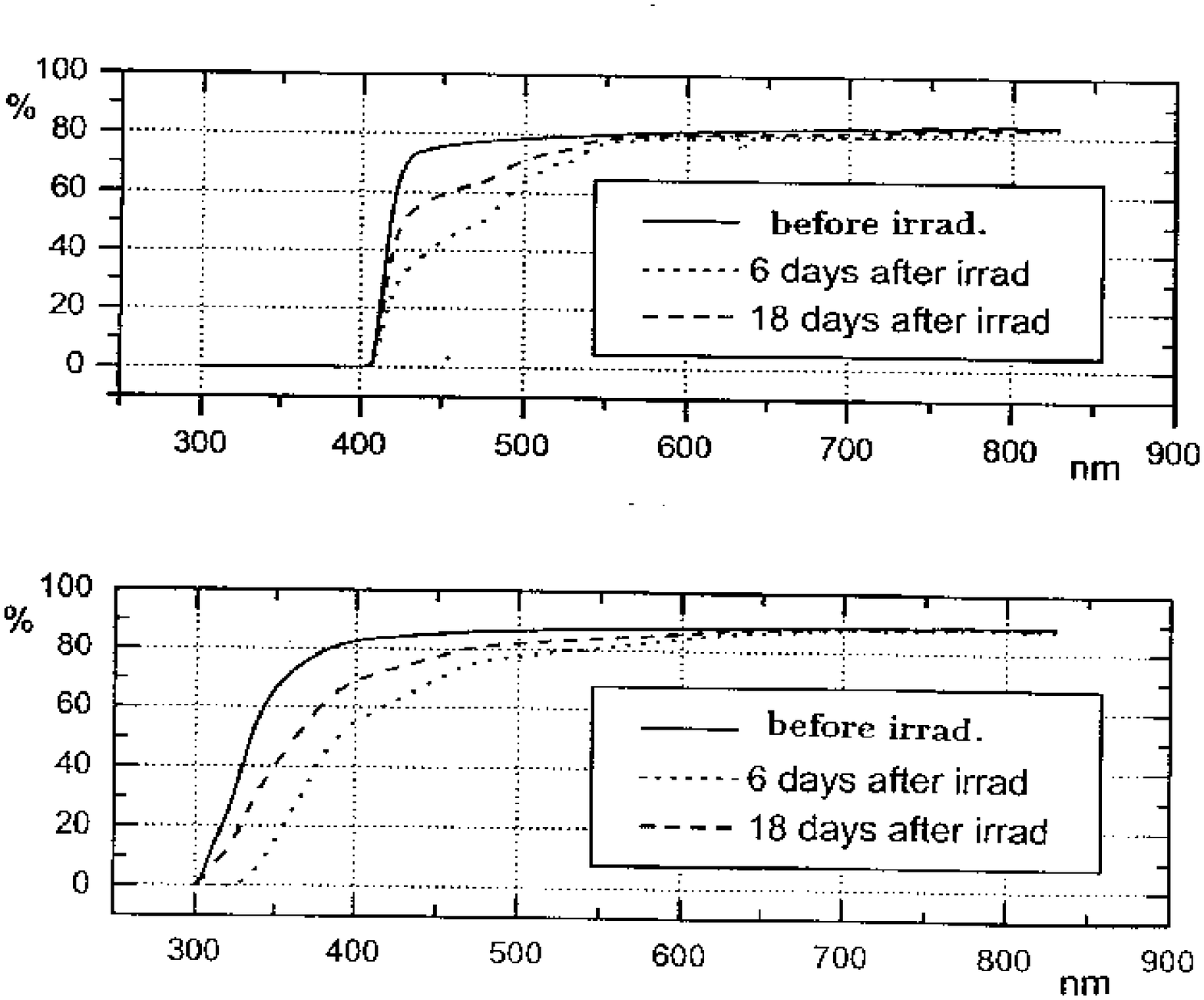,height=100mm,width=155mm,%
    bbllx=0pt,bblly=130pt,bburx=620pt,bbury=670pt,clip=,angle=0}
\end{picture} 
\caption{Optical transission of (a) scintillator and (b) plexiglass
in the wavelength region $300 \leq \lambda \leq 800\,$nm.}
\label{fig:trans}
\end{figure}
The samples under study and several dosimeters
were placed on the collimator of the bremsstrahlung
beam arising from the collision of the accelerated electrons with 
an internal target. After the exposure, the dosimeters were used to 
determine the dose delivered.
The most important property 
studied was the change of the fibre attenuation length;
the mean free path of photons in the fibre.
Using the procedure described in~\cite{BARAN}, investigations 
were performed with BICRON fibres which 
had suffered a dose of $1\,$Mrad.
The measurements show that the photon mean free path in the
fibre decreased from $\sim 130\,$cm to $\sim 60\,$cm after
irradiation. As the expected yearly doses resulting from 
luminosity running
at the position of the ET-8 are about a factor of 
ten less than this,
this effect is not large.

The scintillator and the plexiglass used in
the light guides of the ET-8 
were also studied. These were irradiated for about
40 hours and received a total dose of $2\,$Mrad. The
scintillator sample then apppeared yellow in colour and the 
plexiglas brown, an effect which 
decreased with time after irradiation.
The optical transparency of
the samples before and after irradiation was 
measured using a  spectrophotometer in the
wavelength range $300$ to $800\,$nm. The results are shown
in figure~\ref{fig:trans}.
The changes due to the irradiation occur essentially in the
short wavelength region. A slow recovery of transparency 
with time is
also visible. Several days after the irradiation the optical
properties were partly restored. 
  
It should be noted that the total decrease in light yield
due to the radiation damage of the active elements of the ET-8, the 
fibres, 
scintillators and plexiglass, does not influence the detector response
or resolution significantly since the detector is continuously
calibrated, as are all the detectors comprising the H1 luminosity system.   
\section{Test results}
\subsection{Event selection and event rates} 
%
\begin{figure}[h]
\centering
\begin{picture}(160,100)(0,-95)
\epsfig{figure=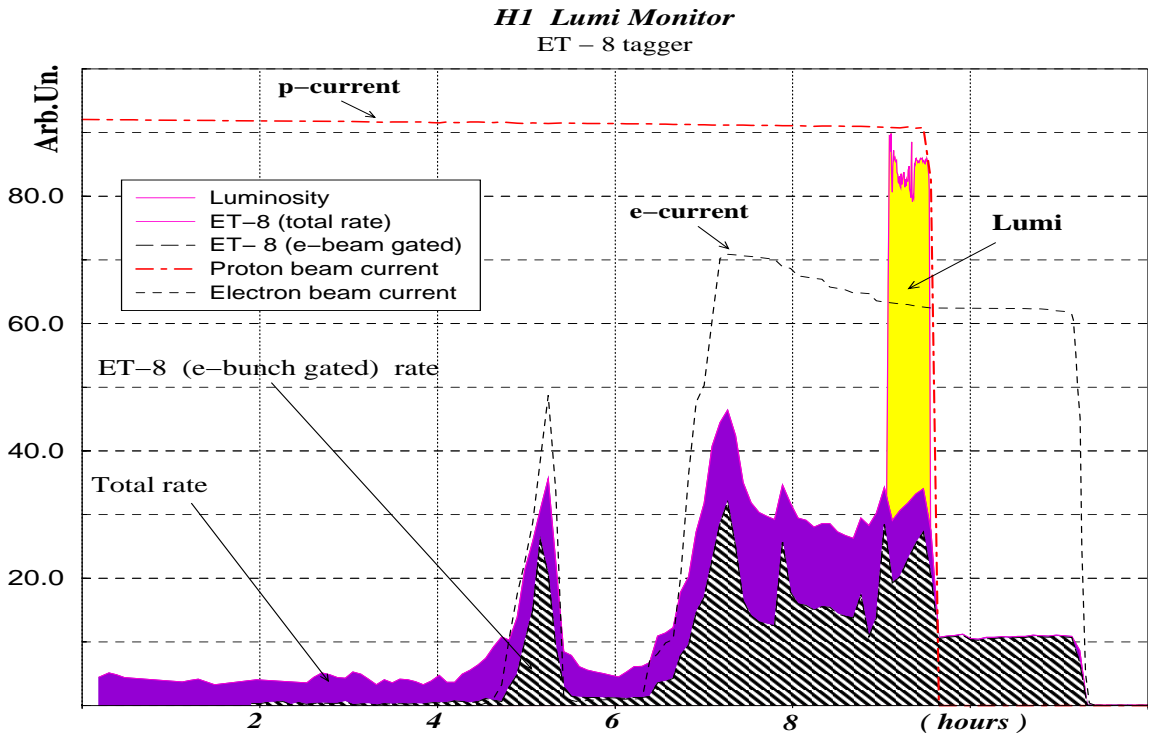,width=100mm,height=155mm,%
    bbllx=20pt,bblly=30pt,bburx=600pt,bbury=700pt,clip=,angle=-90}
\end{picture}
\caption{The event rate in the  ET-8 for a range of beam 
         and luminosity conditions.}
\label{fig:rate}
\end{figure}
%

The 1996 HERA running schedule originally foresaw a long 
shutdown in which it was planned to install a section of 
beampipe with an exit window in front of the ET-8. This 
schedule was modified, the long shutdown did not take place,
so the new beampipe could not be inserted. The
prototype ET-8 calorimeter was thus put into the HERA 
tunnel at a position as close to that foreseen for the 
ET-8 as possible. This was at about $z = -7\,$m, 
so the prototype was dubbed
the ET-7. Test results obtained using the  ET-7 operation 
were presented in~\cite{LIV}.
The installation of the new beam pipe and the 
ET-8 was performed in 1997.  

Figure~\ref{fig:rate} shows, as a function of time $t$, 
the rates observed in the ET-8
during a luminosity run and the beam studies
performed before and after that run.
During the first 2 hours, HERA contained only a stable $920\,$GeV 
proton beam 
and the observed ET-8 rate is a consequence of
the proton beam halo particles hitting
the detector. At $t=2\,$hours, 
a low current electron beam was injected at an energy of $12\,$GeV and
kept at this energy until $t=5\,$hours. 
A large part of the electron beam was lost at $t=5\,$hours 
during attempts to inject a higher electron
current. At $t=7\,$hours, a high current electron beam  
was successfully injected at an energy of $12\,$GeV 
and then ramped up to $27.6\,$GeV. 
During this period the ET-8 rate contains the abovementioned 
proton beam induced component, but also a component
resulting from the presence of the electron beam in
HERA. Both electrons scattered off the residual gas
in the HERA beam pipe and ``off-momentum'' electrons
contribute to this component. 
Luminosity running was established at $t=9\,$ hours.  Soon after  
this the proton beam was
lost, but the electron beam was kept for a further two hours. Hence,
the final part of figure~\ref{fig:rate} shows
the rate induced by the full energy electron beam alone.
The measurements during luminosity running show
clear evidence of an increased 
electron beam gated rate, whereas the proton beam induced component 
of the rate remains approximately constant.
This figure illustrates graphically the problem that 
must be solved by the ET-8 and
other luminosity system triggers. These must identify
with high efficiency the events from 
electron-proton interactions
while discarding as much as possible of the large background.  

\begin{figure}[ht]
\centering
\begin{picture}(150,75)(-5,-70)
\epsfig{file=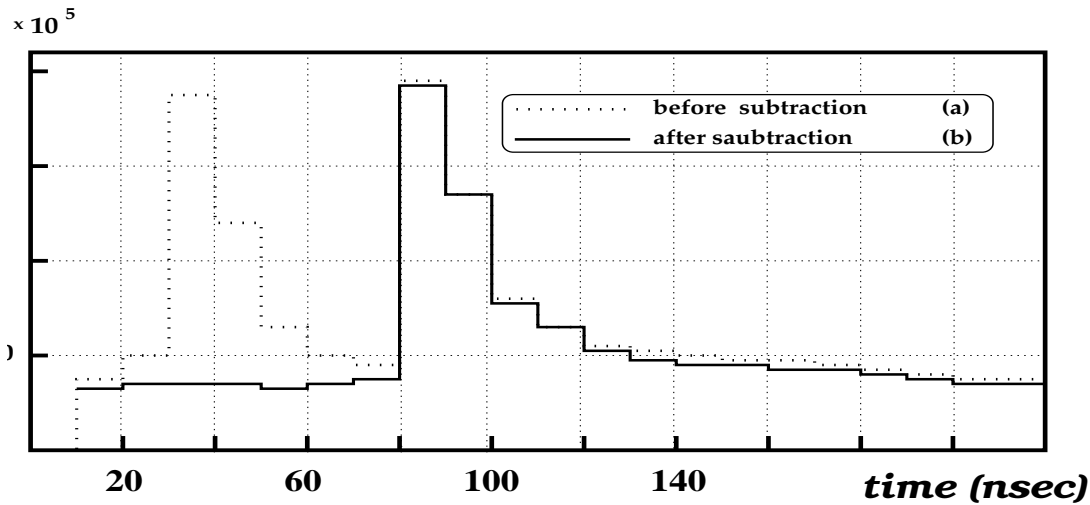,height=145mm,width=100mm,%
    bbllx=0pt,bblly=20pt,bburx=520pt,bbury=640pt,clip=,angle=-90}
\end{picture}
\caption{The averaged FADC's pulse shape for the events triggered by
          {\it H1 - ET-8} subtrigger: 
          {\bf a} - with proton halo and
           {\bf b} - after subtraction proton halo.}
\label{fig:FADC}
\end{figure}

The time structure of the ET-8 signals was measured using
flash analogue to digital converters. 
The averaged FADC pulse shape of the ET-8 response for the
events triggered by the ET-8 trigger is shown in
figure~\ref{fig:FADC}. 
Two spikes are visible in curve (a) on the figure.
The first corresponds to proton halo particles hitting the ET-8, 
the second to the impact of electrons. The time between 
the two is $54\,$nsecs, as expected given the distance 
betwen the interaction point and the ET-8. 
This interpretation is confirmed by the results shown 
in curve (b) which was obtained when signals 
from other detectors were used to subtract events
caused by proton halo particles. 
The timing information visible in 
figure~\ref{fig:FADC} was used 
in the ET-8 trigger, described in more detail below.
   
In order to produce the level one trigger output for the ET-8 detector 
two independent conditions were combined. The first, 
the electron time window condition, was formed by requiring 
that the ET-8 signal occur
within the time interval expected for electron-proton 
interactions. The second, 
the threshold condition, required
that the energy in the ET-8 detector be more than $E_{thr}$. 
During the first stage of ET-8 operation, $E_{thr} \sim 1.0\,$GeV.
\subsection{Detector calibration}
The calibration of the ET-8, and of all the
detectors forming the luminosity system, 
is peformed using the Bethe-Heitler 
process $ep \rightarrow e\gamma p$. Since the energy 
transfer to the proton is negligible, the relation  
$$
   E_{Tagger} + E_{\gamma} = E_e
$$
must hold.
  
The ET-8 calibration procedure relies on the fact 
that the PD has already been calibrated, together 
with the main electron tagger (ET). In addition,
the absolute energy scale of the PD can be determined
using the high energy edge of the bremsstrahlung spectrum~\cite{GOG}.
%
\begin{figure}[ht]
\centering
\begin{picture}(160,70)(0,4)
\epsfig{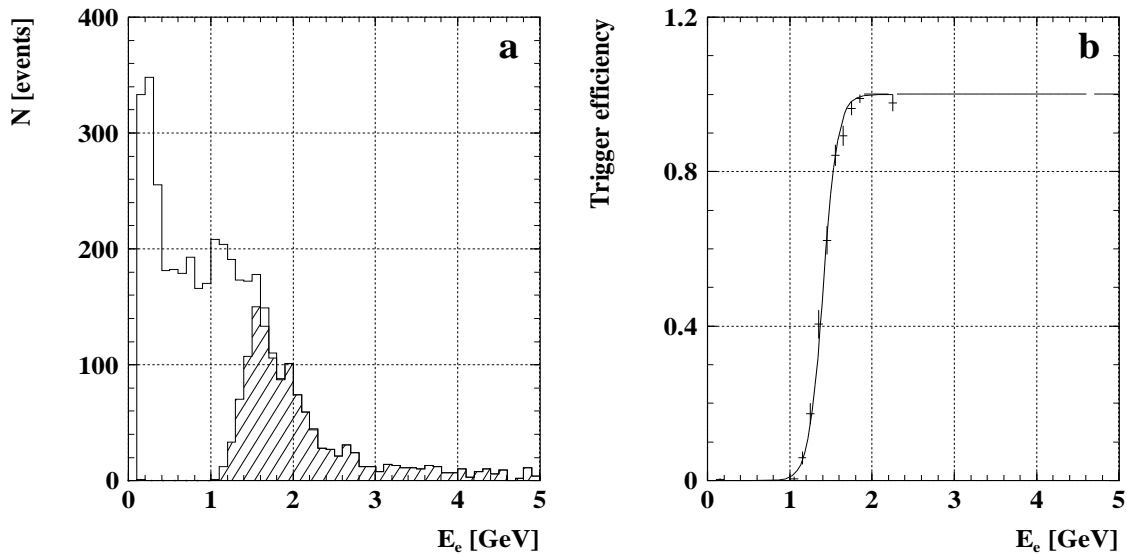}
\end{picture}
\caption{(a) Energy measured in the ET-8 for events 
     which fired the level one ET-8 trigger (shaded histogram) and 
     for a random trigger sample (open histogram);
     (b) efficiency of ET-8 trigger as a function of energy.}
\label{fig:trig}
\end{figure}
%
The first step of the calibration procedure
is to select events with an impact point in 
the front plane of the ET-8 which is well separated from the 
detector edges. 
The second step, described in detail in~\cite{DE24},
determines the calibration coefficients $C(n)$ by minimizing
the following sum:
$$
\sum_{i} \Big(\sum_{n=1}^{N} C(n)\ast A_{i}(n) -E_e\Big)^{2},
$$ 
where $N$ is the total number of
channels, $A_{i}(n)$ is the amplitude of the signal
detected in the $n^{\rm th}$ channel in 
the $i^{\rm th}$ event and 
$C(n)$ is the calibration coefficient for the
$n^{\rm th}$ ET-8 channel.
Using these calibration constants,
the absolute energy spectrum shown in 
figure~\ref{fig:trig}(a) was obtained. The shaded area represents
the energy distribution of the scattered electrons in 
the ET-8 in events in which ET-8 trigger was set. It
has a maximum around $\sim 1.5\,$GeV. 
The influence of the threshold $E_{thr} \sim 1.0\,$GeV
is clearly visible. The open histogram in the same figure contains all
events taken by a random trigger. The background contribution 
peaks strongly at small energies.    
\begin{figure}[ht]
\centering
\begin{picture}(140,75)(0,0)
\epsfig{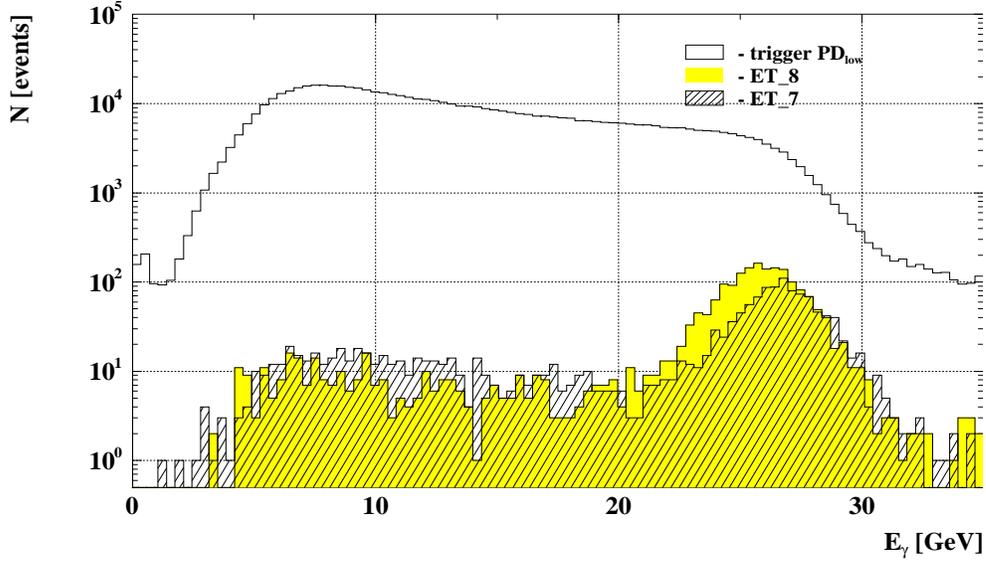}
\end{picture}
\caption{The photon energy spectra measured in the PD;
     the solid line shows that for all measured events, 
     the shaded areas show the spectra for 
     events triggered by the ET-7 and ET-8.}
\label{fig:PDenergy}
\end{figure}
This background arises from off-momentum electrons
in random coincidence 
with a signal from the main H1 detector.
Comparing these two histograms allows the 
determination of the efficiency
of the ET-8 trigger as a function of the energy detected in 
the tagger. This is shown in figure~\ref{fig:trig}(b).

Figure~\ref{fig:PDenergy} shows the measured energy spectrum of 
Bethe-Heitler photons
in the PD. 
%
\begin{figure}[ht]
\centering
\begin{picture}(150,70)(0,0)
\epsfig{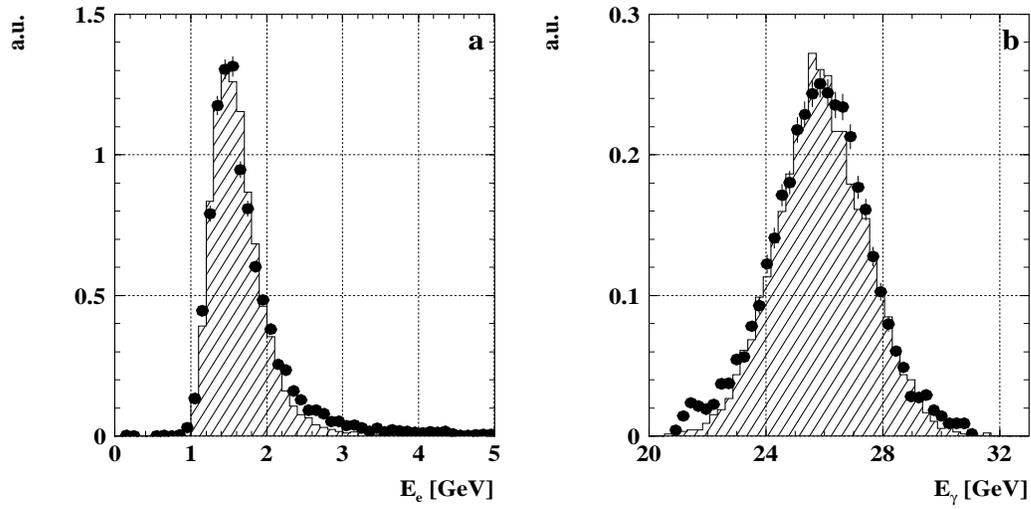}
\end{picture}
\caption{(a) The energy spectrum measured in the ET-8 (points) 
             for bremsstrahlung events with a MC simulation (histogram);
    (b) The measured $E_{\gamma}$ spectrum (points) in
        the PD for bremsstrahlung events in which the electron 
        was detected in the ET-8, the MC prediction is shown
        as a histogram.}
\label{fig:energy}
\end{figure}
The solid line shows all measured events with 
significant energy in the PD. The shaded
areas show the result of demanding coincidences with
trigger signals from the ET-7 or the ET-8.
\begin{figure}[ht]
\centering
\begin{picture}(160,80)(0,1)
\put(70,75){\large {\bf (a)}}
\put(140,75){\large{\bf (b)}}
\epsfig{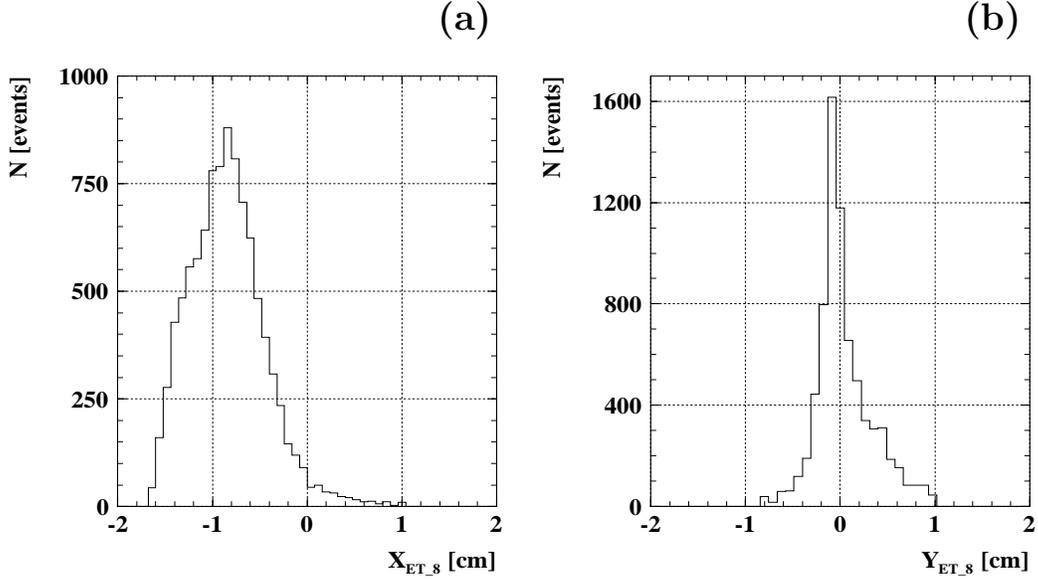}
\end{picture}
\caption{Distribution of the position of electrons scattered into
         the ET-8:
   (a) $x$ co-ordinate of the detected particles at the tagger front
      plane;
   (b) $y$ co-ordinate in the same plane.}
\label{fig:hitpos}
\end{figure}
%
\begin{figure}[ht]
\centering
\begin{picture}(120,80)(-15,0)
\epsfig{file=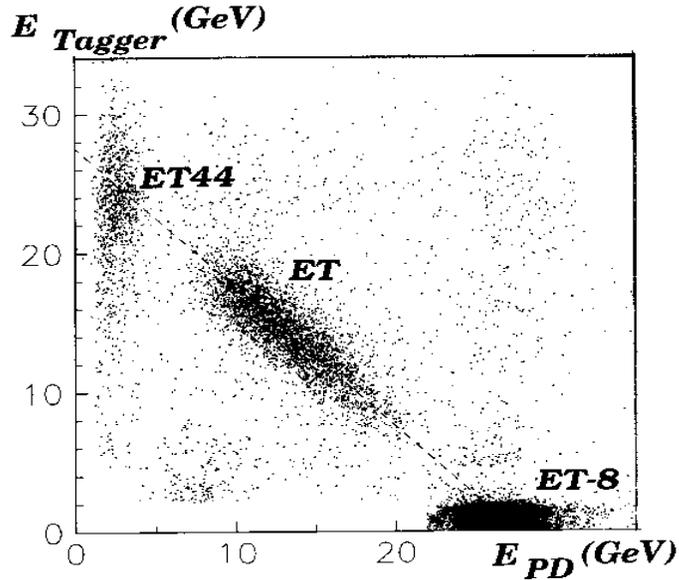,height=80mm,width=90mm,%
    bbllx=65pt,bblly=150pt,bburx=550pt,bbury=650pt,clip=,angle=0}
\end{picture}
\caption{The $E_{\gamma}$ - $E_{e}$ correlation for bremsstrahlung
     events detected by the H1 luminosity system for 
     the ET33, ET44 and ET-8.}
\label{fig:Ecorr}
\end{figure} 
The mean energies of these latter distributions are about $26.2\,$GeV
for the ET-7 and $25.8\,$GeV for the ET-8, in agreement 
with the results of MC simulations for the average 
energy of bremsstrahlung photons inside the corresponding tagger
acceptance.

The agreement between data and simulations is 
further demonstrated in figure~\ref{fig:energy},
in which the measured energy spectrum in the ET-8 tagger 
and the PD-detector for Bethe-Heitler events in which the
electron was detected in the ET-8 are 
shown with the Monte Carlo predictions.
The simulation includes the effects of
detector resolution, trigger efficiency and pile-up.

The horizontal and vertical coordinates of the points 
at which electrons hit 
the front plane of the tagger
are determined by comparing  
the energies deposited 
in the various modules of the ET-8. The 
reconstructed $x$ and $y$ co-ordinate 
distributions 
for a sample of data are presented
in figure~\ref{fig:hitpos}.
 
Figure~\ref{fig:Ecorr} illustrates the 
correlation between
the reconstructed electron energies
$E_{Tagger}$ and photon energies $E_{\gamma}$ 
for bremsstrahlung events.
These were selected by requiring a coincidence 
between the PD trigger and any of the electron tagger triggers.
Due to the large difference in the trigger rates of
the different taggers
(ET44 $\simeq 280\,$kHz, ET $\simeq 170\,$kHz and
ET-8 $\simeq 15\,$kHz at a luminosity of $5 \times 
10^{30}\,$cm$^{-2}$s$^{-1}$) these data were taken
with differing pre-scale factors for the 
various tagger triggers.
The distributions
of $E_{Tagger}$ + $E_{\gamma}$ are 
well described 
by a Gaussian function with the expected 
mean value of $27.6\,$GeV, the electron beam energy, 
and a standard deviation of $ \simeq 1.3\,$GeV. 

\begin{figure}[h]
\centering
\begin{picture}(150,90)(0,0)
\epsfig{file=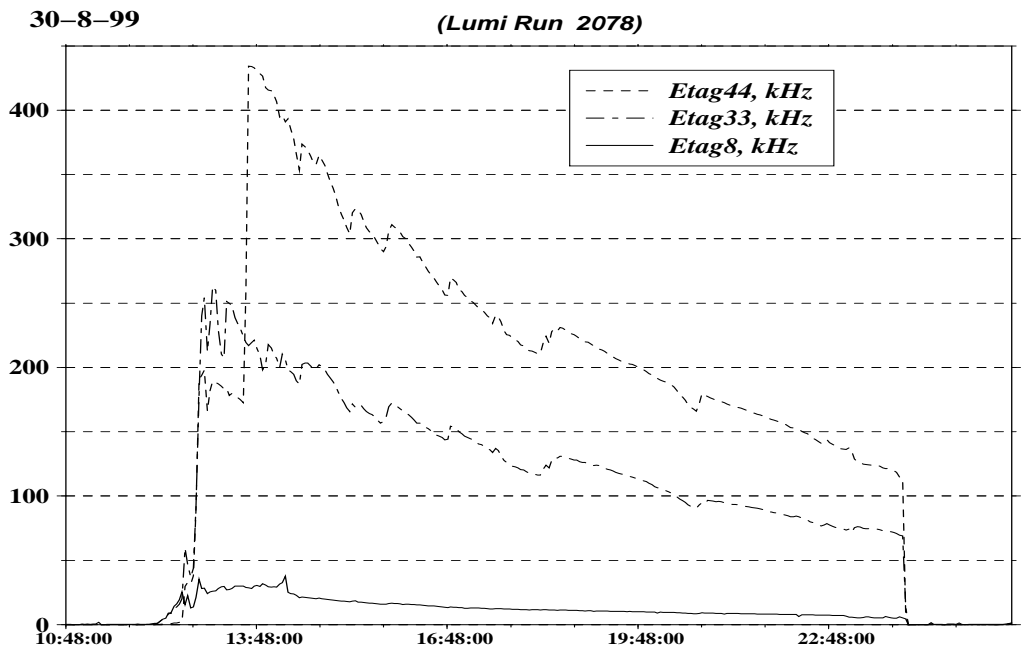,height=90mm,width=140mm,%
    bbllx=1pt,bblly=1pt,bburx=580pt,bbury=478pt,clip=,angle=0}
\end{picture}
\caption{The behaviour of the rates measured in the various 
electron taggers during one HERA fill.}
\label{fig:ratecorr}
\end{figure}

The event rates recorded by the different taggers
within one electron fill (one luminosity run)
are presented in figure~\ref{fig:ratecorr}. 
The observed rates in the various taggers are in good
agreement with
Monte Carlo simulations of 
bremsstrahlung events, taking into account the acceptance of the 
taggers.  Note the sharp rise in the 
ET44 rate at the time the detector
was moved into its working position close to the electron beam axis. 
\subsection{Physics results}
\begin{figure}[h]
\centering
\begin{picture}(120,90)(0,0)
\epsfig{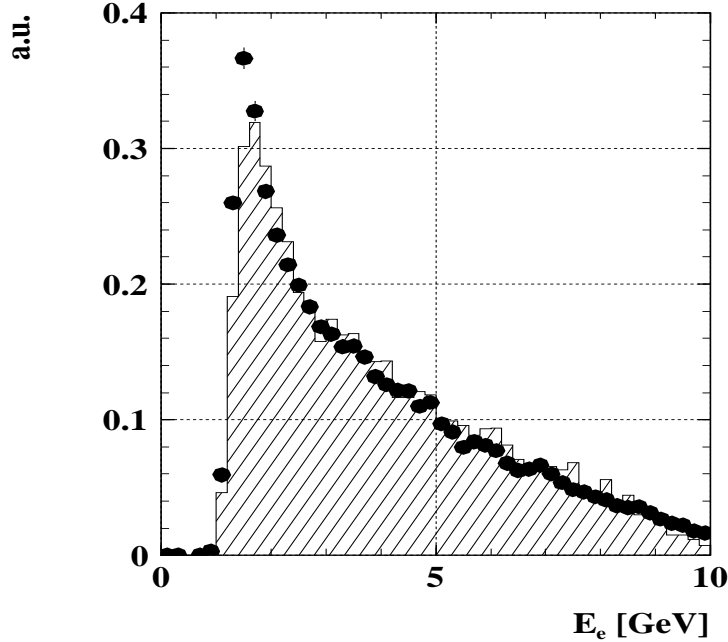}
\end{picture}
\caption{Electron energy spectra (points) measured with the 
         ET-8 for photoproduction
         events, the histogram shows the MC prediction.}
\label{fig:genergy}
\end{figure}

The incorporation of the ET-8 into the H1 detector enables the
identification of photoproduction events at high $W_{\gamma p}$
and allows the accurate
reconstruction of the interacting photon's energy. 
The energy distribution of the scattered
electrons for candidate photoproduction events 
is shown in
figure~\ref{fig:genergy}. 
The MC prediction for this 
distribution is shown in
the same figure. The Monte Carlo  
describes the data well. 
The data obtained using the ET-8 provide the
basis for the study of many physics topics.
As an illustration of the potential of the device, a
preliminary measurement of the charged particle multiplicity in 
photon-proton interactions is presented here.
This was 
made using minimum bias data collected by H1 in the autumn of 1997.

To study the energy dependence of the charged particle multiplicity
in $\gamma p$ interactions, three data samples triggered by 
the three 
electron taggers were used. This provides a measurement in
three $W_{\gamma p}$ regions, determined by the 
acceptance of the taggers, with average values of 
$\langle W \rangle = 100, 200$ and $290\,$GeV.

\begin{figure}[ht] \centering
\begin{picture}(125,95)(0,1)
\epsfig{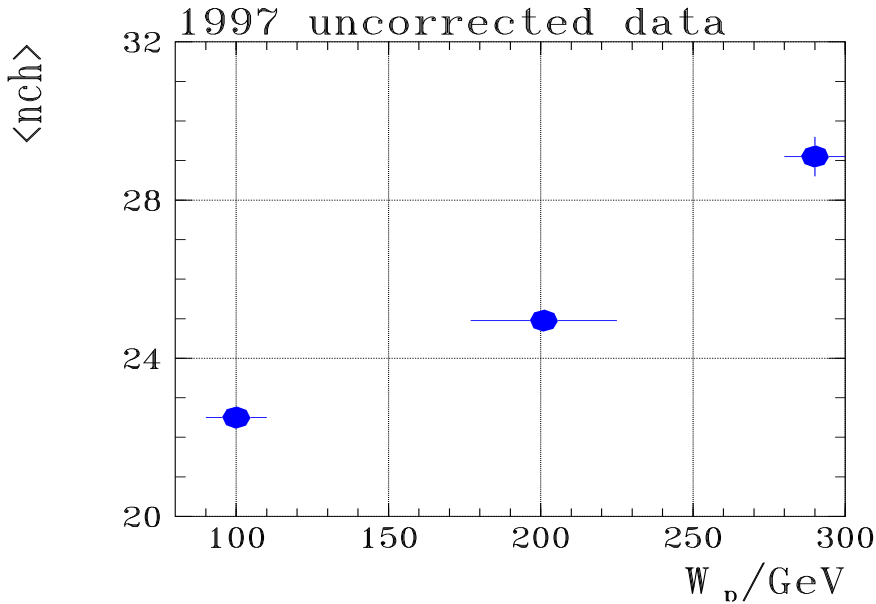}
\end{picture}
\caption{Energy dependence of the charged particle multiplicity
in photoproduction.}
\label{fig:mult}
\end{figure}

The acceptances of all the taggers  
were determined from the data, using events from the 
Bethe-Heitler reaction  $ep \rightarrow e \gamma p$ and the
method described in~\cite{H1COL}. The 
average charged particle multiplicity as a function of 
$W_{\gamma p}$ is
shown in figure~\ref{fig:mult}. 
Only statistical errors are given as the data are not 
corrected for the H1 tracker acceptance and efficiency.

The following criteria were applied in the analysis:
\begin{itemize}
  \item  Good quality runs were selected in which 
         all the main components of H1 were fully operational,
         the central and forward trackers, 
         the LAr and Spacal calorimeters, 
         the electron taggers and the ToF system. 
  \item  The reconstructed event $z$-vertex was required 
         to be within $30\,$cm
         of the nominal interaction point.
  \item  A cut on $y_{JB}$, measured using the main H1 detector,
         was used to remove proton beam gas events
         as described in~\cite{H1COL}. 
\end{itemize}
The level of electron beam gas contamination 
($2$ to $6\%$)  was determined from pilot bunch studies and 
statistically subtracted.

To determine the event multiplicity, all vertex fitted charged tracks
with $p_T > 0.15 GeV/c$ lying in the polar angular
interval $15^{\circ} < \theta < 155^{\circ}$ in the 
laboratory system were used.

\section{Conclusions}
Following the investigation of various possible technical 
solutions, an extremely compact spaghetti type 
electromagnetic calorimeter, the ET-8, was built
and installed within the H1 detector. 
This calorimeter makes possible the detection 
of electrons scattered in photoproduction interactions 
in which the photon takes a large 
proportion of the initial electron energy. 
Acceptance studies showed that 
the optimum position for the calorimeter, given the 
constraints imposed by the HERA machine, 
was at about $8\,$m from the interaction point
in the electron direction.
Studies of the materials used showed that they were
able to withstand the high radiation doses to which they 
are subjected in their position close to the HERA beams.
Tests of both a prototype calorimeter, 
the ET-8 and the associated trigger, reconstruction 
and calibration schemes have revealed that they function 
as desired. First preliminary physics results have been obtained.
\section{Acknowledgements}

We gratefully acknowledge the support of the DESY directorate 
and the machine group of the LPI synchrotron 
(Troitsk). Those of us from outside DESY wish to thank the 
DESY directorate for the kind hospitality extended to us. 
It is also a pleasure to thank  
P. Biddulph and R. Eichler for active support and helpful discussions.
We would like to thank INTAS, RFBR (grant INTAS-RFBR 95-0679) 
and the UK's PPARC for financial support.

\end{document}